\documentstyle{mn}
\def\MSUN{\rm M_{\odot}}

\def\MSUNYR{\rm M_{\odot}\,yr^{-1}}
\def\MDOT{\dot{M}}

\newbox\grsign \setbox\grsign=\hbox{$>$} \newdimen\grdimen \grdimen=\ht\grsign
\newbox\simlessbox \newbox\simgreatbox
\setbox\simgreatbox=\hbox{\raise.5ex\hbox{$>$}\llap
     {\lower.5ex\hbox{$\sim$}}}\ht1=\grdimen\dp1=0pt
\setbox\simlessbox=\hbox{\raise.5ex\hbox{$<$}\llap
     {\lower.5ex\hbox{$\sim$}}}\ht2=\grdimen\dp2=0pt
\def\simgreat{\mathrel{\copy\simgreatbox}}
\def\simless{\mathrel{\copy\simlessbox}}

\title[Line-driven disk wind models]{ Line-driven disk wind models with an 
improved line force}

\author[D. Proga et al]{ Daniel Proga$^{a,b}$, James M. Stone$^c$, and
Janet E. Drew$^a$ \\
$\rm ^a$ Imperial College of Science, Technology and Medicine, 
Blackett Laboratory, Prince Consort Road, London SW7 2BZ, UK \\
$\rm ^b$ {\bf new address} 
Laboratory for High Energy Astrophysics, NASA Goddard Space Flight Center,
                             Greenbelt, MD 20771, USA\\
$\rm ^c$ Department of Astronomy, University of Maryland, College Park 
MD~20742, USA\\
E-mail: d.proga@ic.ac.uk, jstone@astro.umd.edu, and j.drew@ic.ac.uk
} 

\begin{document}
\maketitle

\begin{abstract}
We describe an efficient method of calculating the radiation pressure 
due to spectral lines, including all the terms in the velocity 
gradient tensor. We apply this method to calculate the two-dimensional,
time-dependent structure of winds from luminous disks.
Qualitative features of our new models are very similar to those we calculated 
including only the dominant terms in the tensor 
(Proga, Stone \& Drew 1998, hereafter PSD). 
In particular, we find that models which displayed 
unsteady  behaviour in PSD are also unsteady with the new method, and gross
properties of the winds, such as mass-loss rate and characteristic velocity
are not changed by the more accurate approach. The largest change
caused by the new method is in the disk-wind opening angle: winds
driven only by the disk radiation are more polar with the new method
while winds driven by the disk and central object radiation
are typically more equatorial.  In the closing discussion, we provide
further insight into the way the geometry of the radiation field and
consequent flow determines the time properties of the flow.

\end{abstract}


\section{Introduction}

There has long been an awareness that radiation pressure due to
spectral lines should be capable of driving winds from luminous disks
(e.g., Vitello \& Shlosman 1988, Murray et al. 1995).  However, the
geometry of the case demands a multi-dimensional treatment that has
only recently been undertaken by Pereyra, Kallman \& Blondin 1997) and
ourselves (Proga, Stone \& Drew 1998, hereafter PSD).  A surprising
outcome of our models was that, in cases where the driving radiation
field is dominated by the contribution from the disk, the flow is
unsteady.  Despite the complex structure of the disk wind in this case,
the time-averaged mass loss rate and terminal velocity fit onto the
same scaling with luminosity as do steady flows obtained where the
radiation is dominated by the central object.  In fact these relations
have been shown to be similar to those well-established by analysis for
spherically-symmetric stellar winds (Proga 1999).

Our models adopt the method for calculating the line acceleration for
one-dimensional radial flows, first introduced by Castor, Abbott \&
Klein (1975, hereafter CAK), that has since been further developed
within the context of stellar winds from hot, luminous OB stars (Friend
\& Abbott 1986; Pauldrach, Puls \& Kudritzki 1986).   In order to
extend the CAK method to describe multi-dimensional disk winds, it is
necessary to accommodate the effects of the three-dimensional velocity
field and the direction-dependent intensity.  These effects can lead to
qualitatively different results compared to those obtained from a
one-dimensional treatment (see, for example, Owocki, Cranmer \& Gayley,
1996, hereafter OCG, on the case of a rapidly rotating star).

The most difficult aspect of calculating the line force in a disk wind
is in the evaluation of the integral involving the velocity gradient
tensor (Q, which controls the anisotropic line opacity) over the entire
solid angle occupied by radiating surfaces.  In our previous models, we
followed the example of Icke's (1980) earlier numerical work on disk
winds driven by continuum radiation pressure.  In this approach, the
integral was evaluated using an angle-adaptive quadrature to ensure an
accurate result.  However, computational limitations required that we
simplify the integrand, retaining only the dominant terms in the
velocity gradient tensor.  Specifically, we kept only the radial
gradient in the evaluation of the acceleration due to the radiation
from the central object, and the vertical gradient for the acceleration
due to the disk.  Moreover, we dropped azimuthal terms that are present
in a rotating flow.  In this paper, we introduce a new quadrature that
avoids any simplification of the integrand.  This allows us to evaluate
the radiation force for completely arbitrary velocity fields within the
context of the CAK formalism.  We can now explore the consequences of
the approximations in our previous work and move onto a more general
formulation.

In this paper we recalculate several disk wind models first presented
in PSD.  We concentrate on assessing how gross properties -- such as
the mass-loss rate, velocity,  opening angle and time behaviour of disk
winds -- change when Q, the velocity gradient tensor, is treated in
full.  We describe our `full-Q' method for a general three dimensional
case in Section~2.  Our new method of numerical evaluation of
the line force is described in Section~3. We show our new results
and compare them with those in PSD  in Section~4,
and discuss the implications in Section~5.

\section{Method} 

To compute the structure and evolution of a line-driven wind
from a luminous disk, we solve the equations of hydrodynamics
\begin{equation}
   \frac{D\rho}{Dt} + \rho \nabla \cdot {\bf v} = 0,
\end{equation}
\begin{equation}
   \rho \frac{D{\bf v}}{Dt} = - \nabla (\rho c_s^2) + \rho {\bf g}
 + \rho {\bf F}^{rad}
\end{equation}
where $\rho$ is the mass density, ${\bf v}$ the velocity,
${\bf g}$ the gravitational acceleration of the central star, and
${\bf F}^{rad}$ the total radiation force per unit mass.
The gas in the wind is taken to be isothermal with a sound speed $c_s$.

We adopt the same geometry and assumptions to compute the radiation
field from the disk and central star as in PSD.  That is, we consider
the disk to be flat, Keplerian, geometrically-thin and
optically-thick.  The radiation field of the disk is specified by
assuming that the temperature follows the radial profile of the so-called
$\alpha$-disk (Shakura \& Sunyaev 1973), and therefore depends only on
the mass accretion rate in the disk, $\MDOT_a$,  and the mass 
and radius of the central star, $M_\ast$  and $r_\ast$.  In models where 
the central star is also radiant, we take into account the stellar 
irradiation of the disk, assuming that the disk re-emits all absorbed 
energy locally and isotropically.  See PSD and below for further details.

As in PSD, we approximate the radiative acceleration due to lines
(line force, for short) using a modified CAK method.  The line force
at a point W defined by the position vector $\bf r$ is
\begin{equation}
{\bf F}^{rad,l}~({\bf{r}})=~\oint_{\Omega} M(t) 
\left(\hat{n} \frac{\sigma_e I({\bf r},\hat{n}) d\Omega}{c} \right)
\end{equation}
where $I$ is the frequency-integrated continuum intensity in the direction
defined by the unit vector $\hat{n}$, and $\Omega$ is the solid angle
subtended by the disk and star at the point W. 
The term in brackets is the electron-scattering radiation force,
$\sigma_e$ is  the mass-scattering coefficient for free electrons,
and $M(t)$ is the force multiplier -- the numerical factor which
parameterises by how much spectral lines increase the scattering
coefficient. In the Sobolev approximation, $M(t)$ is a function
of the optical depth parameter
\begin{equation}
t~=~\frac{\sigma_e \rho v_{th}}
{ \left| dv_l/dl \right|},
\end{equation}
where $v_{th}$ is the thermal velocity, 
and $\frac{dv_l}{dl}$ is the velocity gradient along the line of sight.

We adopt the CAK  analytical expression
for the force multiplier as modified by Owocki, Castor \& Rybicki 
(1988, see also PSD)
\begin{equation}
M(t)~=~k t^{-\alpha}~ 
\left[ \frac{(1+\tau_{max})^{(1-\alpha)}-1} {\tau_{max}^{(1-\alpha)}} \right]
\end{equation}
where k is proportional to the total number of lines,
$\alpha$ is the ratio of optically thick to optically-thin lines,
$\tau_{max}=t\eta_{max}$ and $\eta_{max}$ is a parameter 
related to the opacity of the most optically thick lines.
The term in the square brackets is the Owocki, Castor \& Rybicki correction
for the saturation of $M(t)$ as the wind becomes optically thin
even in the strongest lines, i.e., 
\begin{displaymath}
\lim_{\tau_{max} \rightarrow 0} M(t)~=~M_{max}~=~
k(1-\alpha)\eta_{max}^\alpha.
\end{displaymath}

In the generalized Sobolev method
$\frac{dv_l}{dl}$ may be written as
$\frac{dv_l}{dl}~=~\hat{n}\cdot{\bf\nabla}(\hat{n}\cdot{\bf v})$,
or as in Rybicki \& Hummer (1978) (see also PSD)
\begin{equation}
\frac{dv_l}{dl}~=~
Q~\equiv~ \sum_{i,j}\frac{1}{2}\left(\frac{\partial v_i}{\partial r_j}
+\frac{\partial v_j}{\partial r_i}\right)n_in_j=\sum_{i,j}e_{ij}n_in_j
\end{equation}
where $e_{ij}$ is the symmetric rate-of-strain tensor, and $v_i$,
$r_i$, and $n_i$ are the components of $\bf v$, $\bf r$, and $\hat{n}$
respectively.  In the generalized 3D case, the flow velocity along
$\hat{n}$ may  not be monotonic, resulting in radiative coupling between
distant parts of the flow and making $t$  a  non local quantity.  Here
we do not take into account the non local effects on $t$ but rather
concentrate on taking into account all terms of $Q$.  

As in PSD, we use the ZEUS-2D code to numerically integrate the
hydrodynamical equations 1 and 2.  We describe our numerical algorithm
for evaluating the line force in the next section.  For a rotating
flow, there may be an azimuthal component to the line  force even in
axisymmetry.  In contrast to the approximate method used in PSD, the
full-Q formulation used here allows us to treat the effects of a
non-zero azimuthal component to the radiation force on the wind
self-consistently.  We examine these effects in detail in section 4.

\section{Numerical Evaluation of the Line Force}

The disk line force is a complicated integral in which the dependences
on geometry, the radiation field and local optical depth are not
separable. In practice, this integral must be evaluated over the whole
computational domain at every time step of the hydrodynamical
calculations.  A fast numerical integration scheme is therefore
essential.

We perform our calculations in spherical polar coordinates
$(r,\theta,\phi)$ with $r=0$  at point C, the center of the central
star.  We measure colatitude, $\theta$, from the rotation axis of the
disk and assume axial symmetry about this axis.  Azimuth, $\phi$, is
measured from a plane perpendicular to the disk plane, containing the
point C and a point W above the disk.  We define the position of a wind
point, W, and of a disk point, D, by  the vectors ${\bf r}=(r,\theta,
\phi=0^o)$ and ${\bf r}_D=(r_D, \theta_D=90^o, \phi_D)$, respectively.

To specify directions $\hat{n}$ about W we use a second  spherical
polar coordinate system which has an origin at W, colatitude,
$\theta_1$, measured from the direction C toward W, and azimuth,
$\phi_1$, measured from a plane perpendicular to the disk plane, and
containing  C and W.  The components  of $\hat{n}$  in  the
$(r,\theta,\phi)$ system expressed in terms of $\theta_1$ and $\phi_1$
are
\begin{equation}
\hat{n} = (n_r, n_\theta, n_\phi)~=~
(~\cos~\theta_1, \sin~\theta_1~\cos~\phi_1, \sin~\theta_1~\sin~\phi_1). 
\end{equation}
and an element of the solid angle is
\begin{equation}
 d\Omega=  \sin \theta_1 d\theta_1 d \phi_1. 
\end{equation}
Note the transformation between this second coordinate system
(on which we evaluate the radiative acceleration) and the original
(which defines the hydrodynamical grid) is a combination of a displacement
of the origin by ${\bf r}$ and rotation by $\theta$.

In the disk plane at $r=r_{D}$, $I({\bf r}, \hat{n})$ is the local
isotropic disk intensity:
\begin{eqnarray}
\rule{0in}{3.0ex}
I_{D}(r_D) & =~\frac{3 G M_\ast \MDOT_a}{8 \pi^2 r_\ast^3} 
\left\{ \rule{0in}{3.0ex}  \frac{r_\ast^3}{r_D^3}\left(1 - 
\left( \frac{r_\ast}{r_D}\right)^{1/2}\right) \nonumber \right.~~~~~~~~~~~~~\\
 & \left. +\frac{x}{3\pi}\left(\arcsin \frac{r_\ast}{r_D} - 
\frac{r_\ast}{r_D} \left(1 - 
\left(\frac{r_\ast}{r_D}\right)^2\right)^{1/2}\right) \right\},
\end{eqnarray}
We include the effects of the irradiation of a disk by  a star for
$x>0$ where $x$ is defined as the ratio between the stellar luminosity 
$L_\ast$ and the disk luminosity, $L_D$ (PSD).
The co-ordinates of ${\bf r}_D$ in the system defining the hydrodynamical
grid, expressed in the second coordinate system centered on W, are: 
\begin{eqnarray}
r_D &=& ({r^2 + d_D^2  -2 r d_D \cos \theta_1})^{1/2}  \\
\theta_D &=& 90^o  \\
\phi_D  &=& \arcsin~\frac{d_D \sin~\theta_1 \sin \phi_1}{({r^2 + d_D^2  -2 r d_D \cos \theta_1})^{1/2} } 
\end{eqnarray}
where 
$
d_D={r \cos~\theta}/({\cos~\theta \cos~\theta_1 -
\sin~\theta~\sin~\theta_1~\cos~\phi_1})
$
is the distance between D and W.
For radiation from the star, the intensity may be written:
\begin{equation} 
I_\ast~=~\frac {L_\ast}{4\pi^2 r_\ast^2}~=~x\frac{G M_\ast \MDOT_a }{8 \pi^2 r^3_\ast}.
\end{equation}
The precise location on the star of the point of emission is not
relevant because we assume that the stellar surface is isothermal.

We split the integration of the line force over $\Omega$ (see eq. 3)
into the integration over the stellar  solid angle, $\Omega_\ast$  and
the disc solid angle, $\Omega_D$.  We take into account the effects due
to shadowing of the disk by the star, and occultation of the star by
the disk, by properly defining the limits of integration for each.  The
contributions to the line force due to the disk and star are
respectively
\begin{equation} 
{\bf F}_D^{rad,e}~=~\oint_{\Omega_D} M(t)
\left({\hat n}\frac{\sigma_e I_D d {\Omega}}{c}\right) 
\end{equation}
and 
\begin{equation} 
{\bf F}_\ast^{rad,e}~=~\oint_{\Omega_\ast} M(t)
\left({\hat n}\frac{\sigma_e I_\ast d {\Omega}}{c}\right).
\end{equation}
The actual variables we use in the integration
are $\mu$ and $\phi_1$ for the radial line force, 
$\mu$ and $\nu_s$ for the latitudinal line force, 
and $\mu$ and $\nu_c$ for the azimuthal line force, where
$\mu=\cos \theta_1$, $\nu_s = \sin \phi_1$
and $\nu_c = \cos \phi_1$. Using these new variables
we can write  $\hat{n}d\Omega$ in eqs (14) and (15) as
\begin{equation} 
\hat{n} d\Omega=(-~\mu d\mu d\phi_1, 
-\sqrt{1-\mu^2} d\mu d\nu_s,
\sqrt{1-\mu^2} d\mu d\nu_c).
\end{equation} 
Note the force multiplier $M(t)$ in equations 14 and 15 depends on the
rate of strain tensor $e_{ij}$ via equations 4 through 6.  The
components of this tensor in spherical polar coordinates are given,
e.g., in Batchelor (1967).  We evaluate these  from the
velocity components on the hydrodynamical grid by
using finite-difference approximations to the terms in $e_{ij}$. 
The ZEUS-2D code
uses a staggered grid such that scalars and the components of vectors
and tensors are centered at different locations on the grid. For example,
discrete values for the density are stored at zone centers, $v_r$ values 
are stored at zone interfaces in the radial direction, and $v_\theta$ 
components are stored at zone interfaces in the latitudinal direction 
(see Stone \& Norman 1992 for details). Subsequently different components
of the line force (and therefore corresponding $M(t)$ and  $e_{ij}$) 
are also defined at different locations on the grid.
To properly evaluate $e_{ij}$ at a given location 
we have to  calculate discrete values of the appropriate variables at,
and variable differences around, this location.
We take simple means of variable pairs if we need to 
calculate a value mid-way between hydrodynamic grid points.

We apply the trapezoidal method to integrate the line force due to both
the star and disk.  We find that  a dozen  quadrature points for $\mu$
and for $\phi_1$ (or $\nu_s$ or $\nu_c$) are sufficient to achieve a
satisfactory accuracy of the disk and stellar line force even for W
near the disk plane and stellar surface.  This represents a
considerable reduction in the number of quadrature points compared to
the original discretization of solid angle used in PSD.  This reduction is 
possible because of the
simple form of the $\hat{n}d\Omega$ factor (see eq. 16) in the full-Q
method used here in comparison to the method implemented in PSD.  For
example, in the disk contribution to the radial component of the line force, 
we here have $n_r d\Omega = -~\mu d\mu d\phi_1$ as compared with the form
in PSD:
\begin{equation}
n_r d\Omega = \frac{r - r_D~\sin~\theta~\cos~\phi_D}{ d_D}
\frac{r \cos\theta}{d_D^{3}} r_D dr_D d\phi_D 
\end{equation}
where $d_D = (r_D^2 + r^2 - 2 r_D r\sin \theta \cos\phi_D)^{1/2}$
(see Appendix~A in PSD).  

To ensure proper cancellation
of the contributions to the net value of the line force from regions of
the disk corresponding to negative (positive) $n_{r}$, we break the
integration over $\mu$ into sub-intervals for negative (positive)
$\mu$.  Similarly, we break the integration over $\phi_1$ into
sub-intervals for negative (positive) $\phi_1$, corresponding to
negative (positive) intervals for both $\nu_s$ or $\nu_c$.

To afford a full recalculation of the line force for all locations at
every time step we need not only a relatively low number of quadrature
points, but also a short computational time per quadrature point.  We
reduce the latter by computing all the terms in brackets in eqs 14 and
15 at the beginning of each model, as they depend only on the radiation 
field geometry. We then use these precalculated factors throughout the 
time evolution. Thus, to evaluate eqs 14 and 15 at every hydrodynamical 
grid point and every time step, we have to (i) calculate $M(t)$, (ii) 
multiply it by the precalculated geometric factors for all quadrature points 
($12^2$ in this case), and (iii) sum up the products.  The price to pay for 
this reduction in computation is an increase in memory because we have to
store all pre-calculated geometric factors for all grid points,
quadrature points and line force components.

\section{Results}

\begin{figure*}
\begin{picture}(180,590)
\put(0,0){\includegraphics{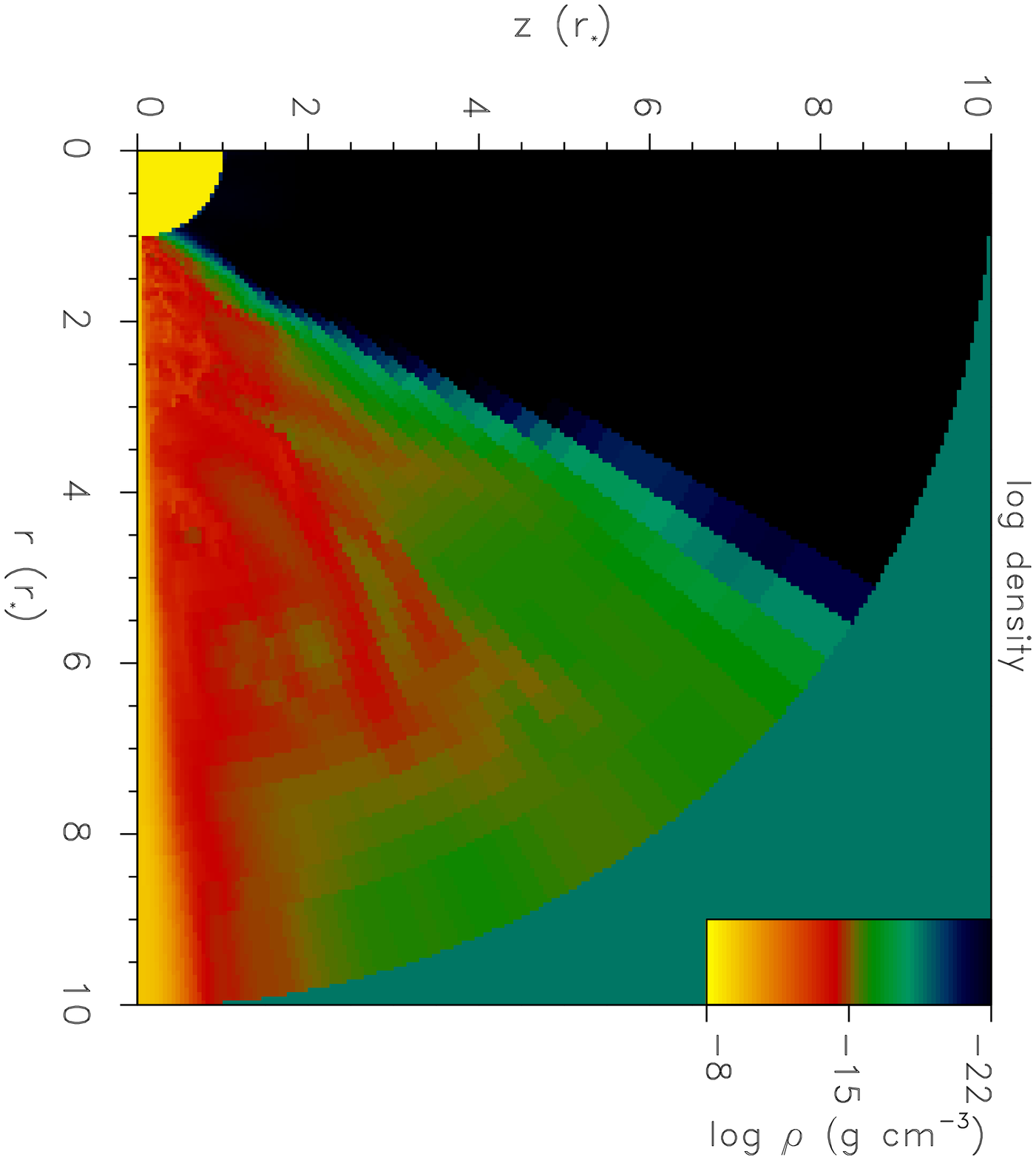}}
\put(0,205){\includegraphics{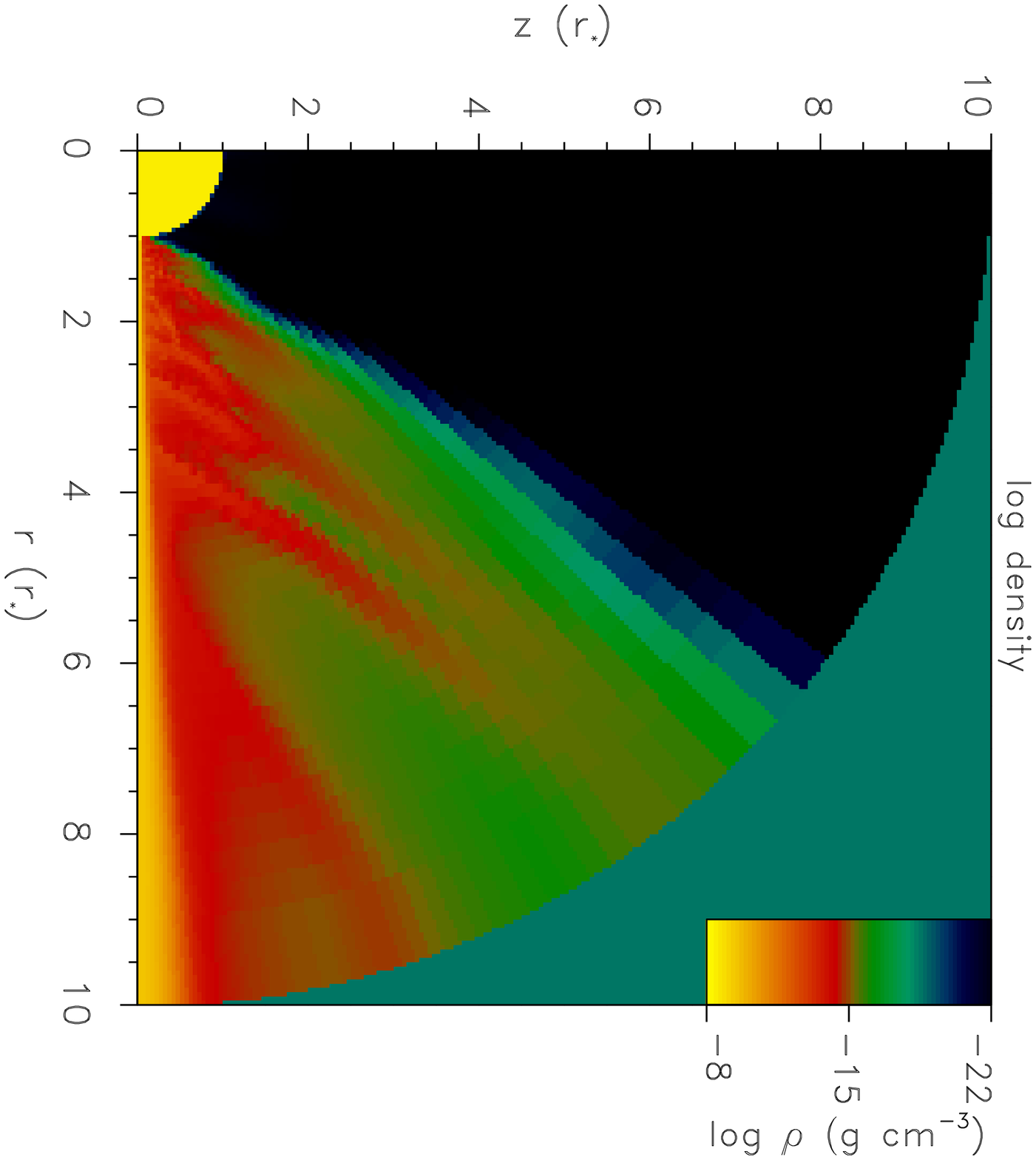}}
\put(0,410){\includegraphics{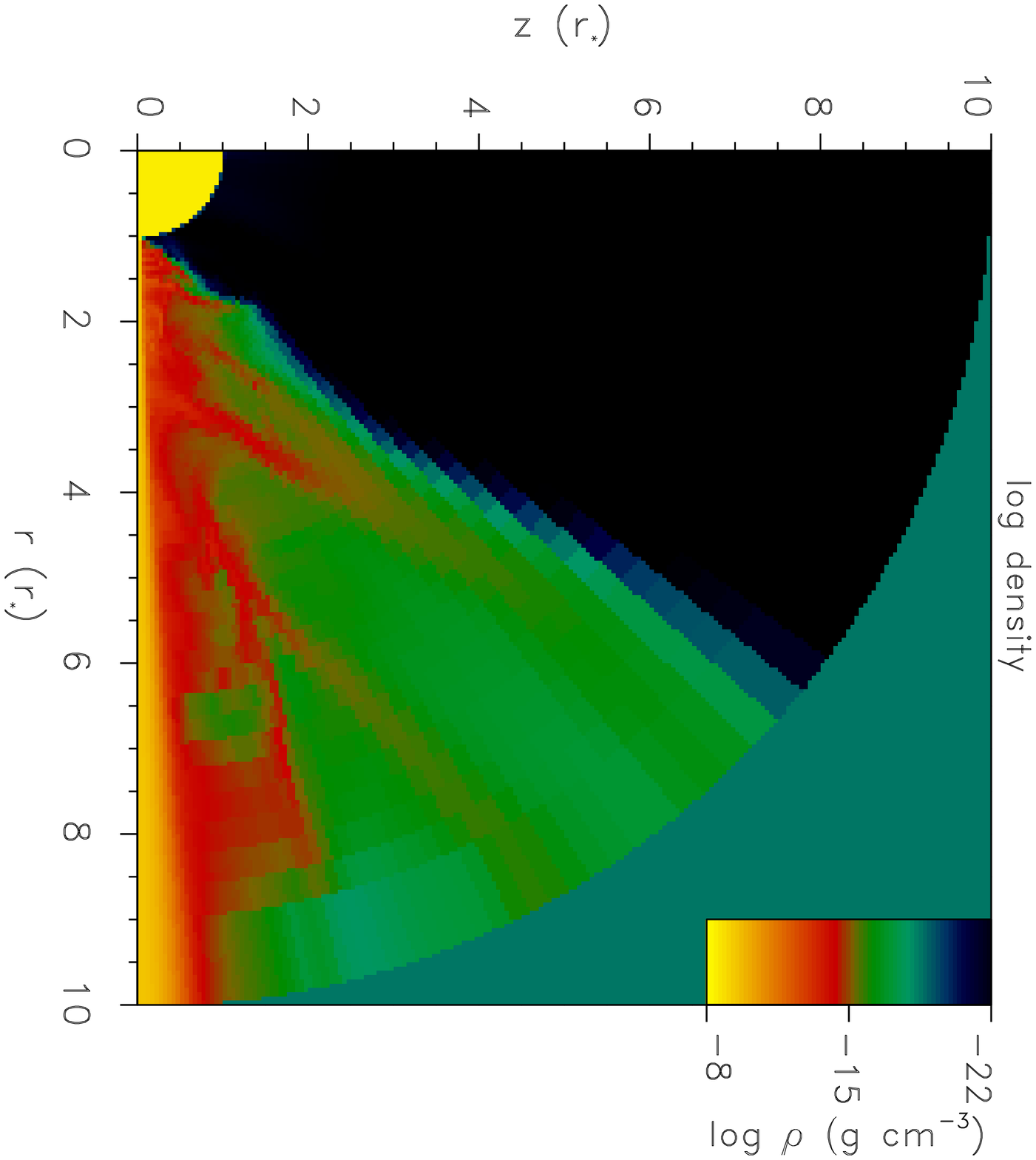}}
\put(90,0){\includegraphics{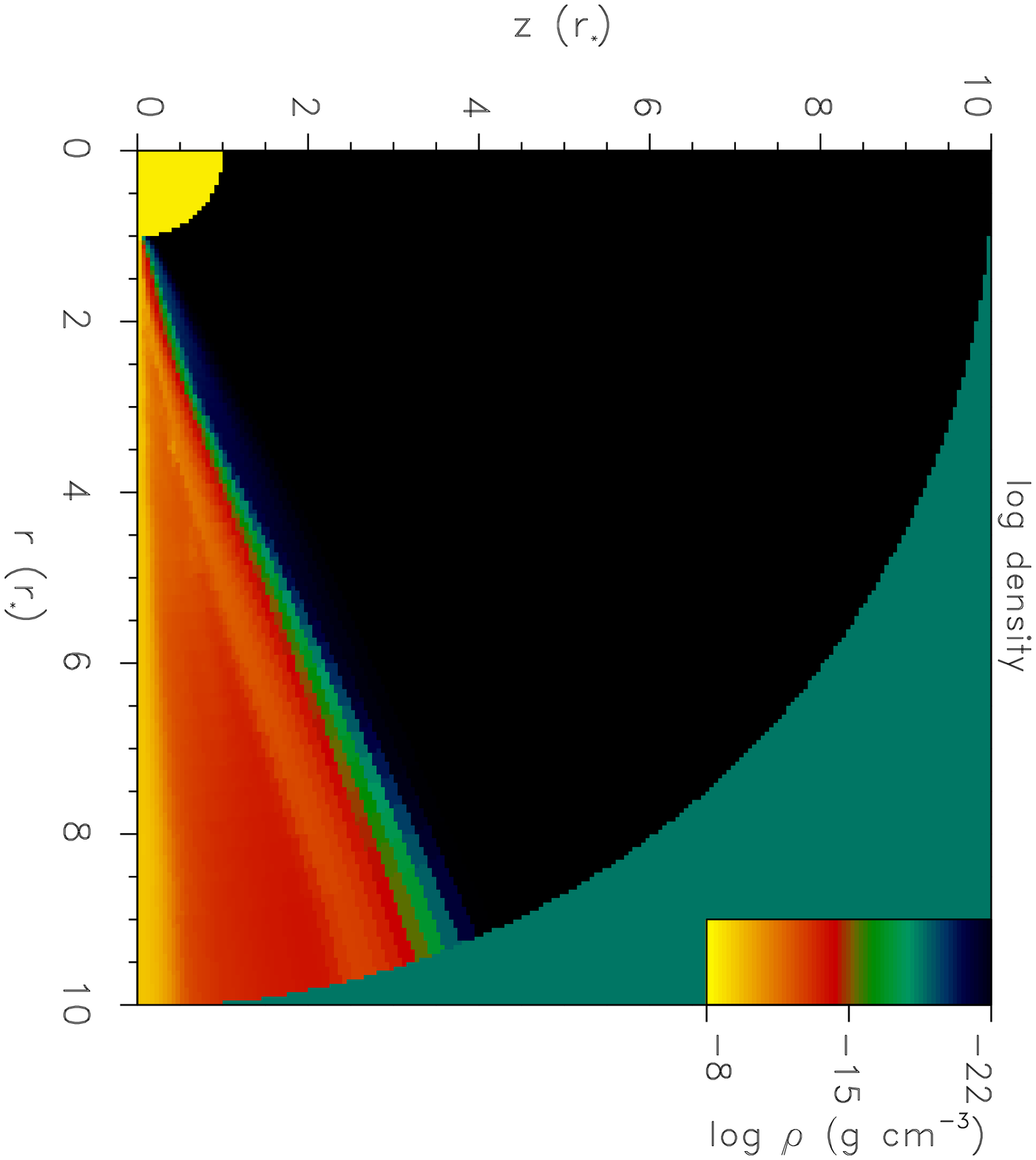}}
\put(90,205){\includegraphics{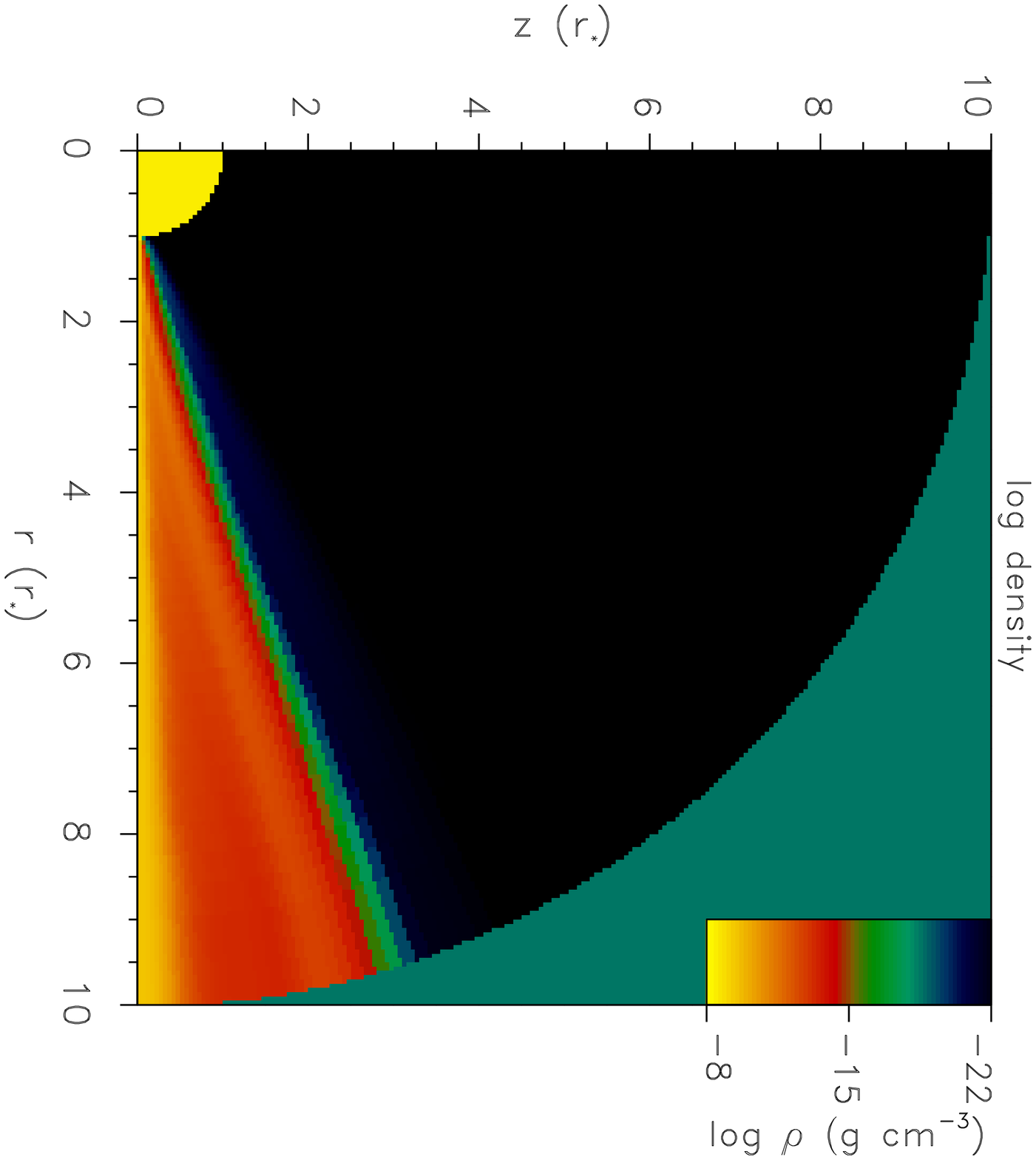}}
\put(90,410){\includegraphics{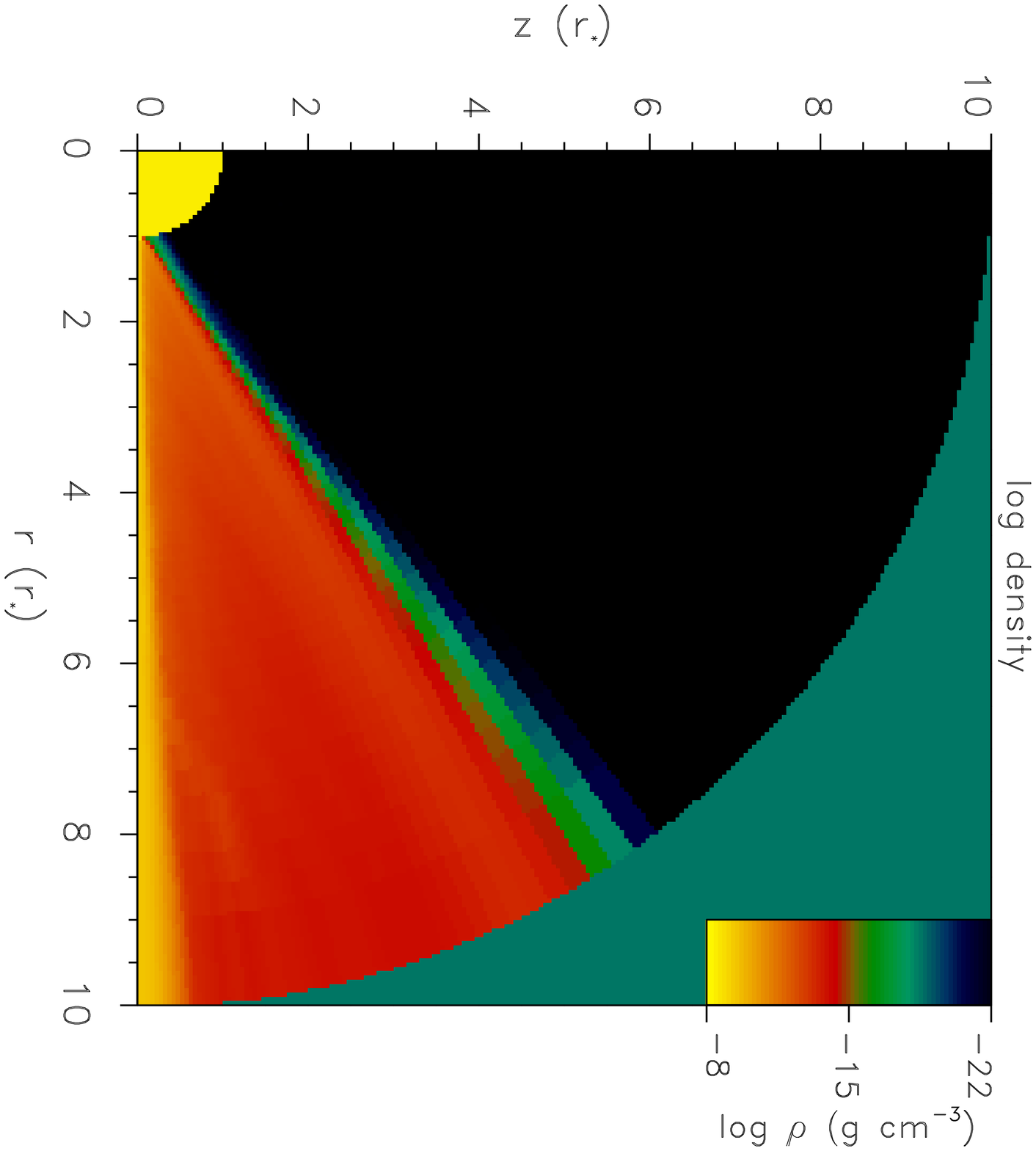}}
\end{picture}
\caption{ 
The density maps in two models computed using the method of PSD (top
panels), the full-Q method but with $e_{r\phi}=e_{\theta\phi}=0$
(middle panels), 
and the full-Q method retaining all terms (bottom panels).
The left column shows the results for the unsteady model~A
while the right column shows results for the steady model~D
(see table~1, and section~4.1 for discussion).}
\end{figure*}

\begin{figure*}
\begin{picture}(180,400)
\put(0,0){\includegraphics{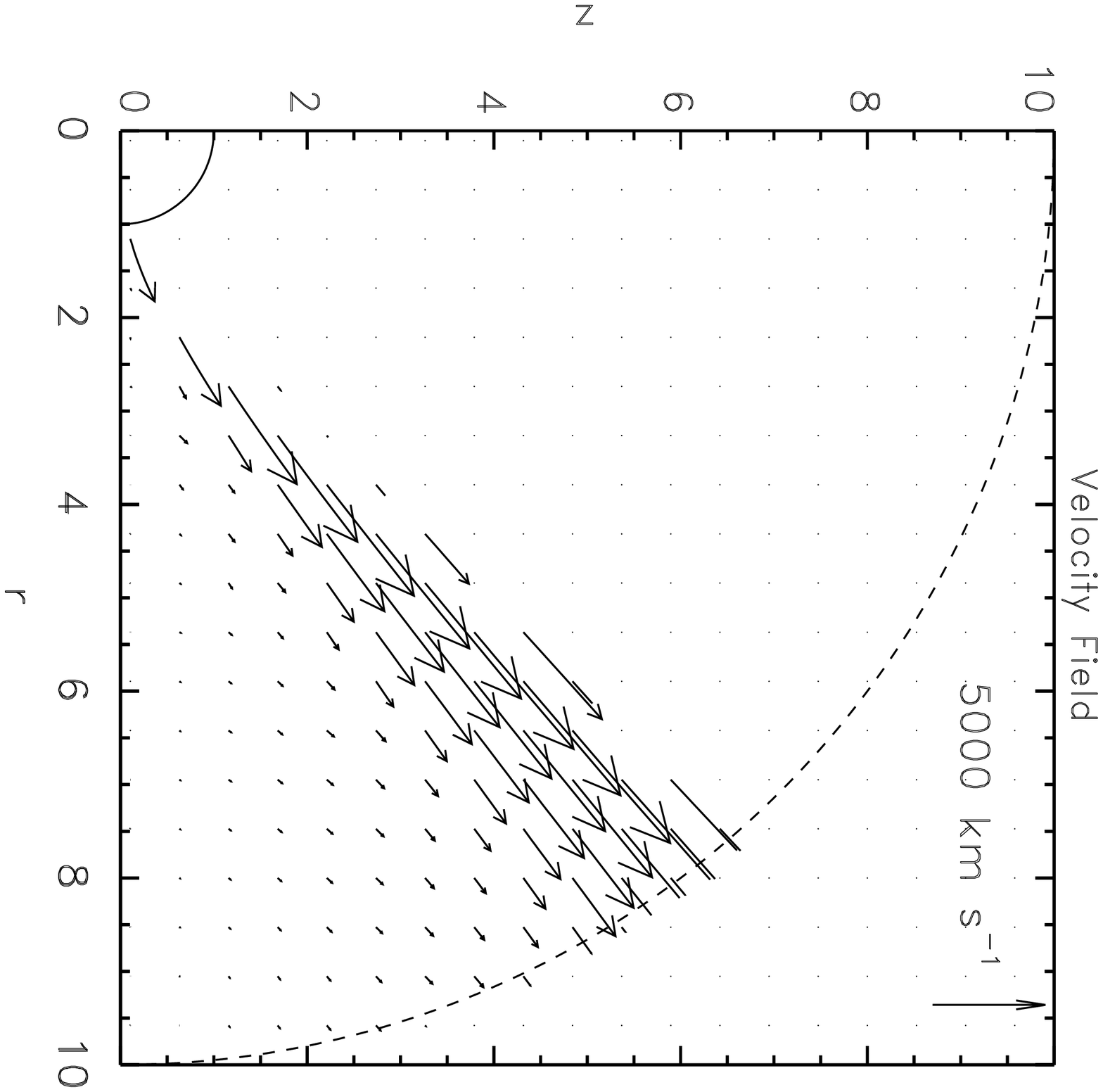}}
\put(0,210){\includegraphics{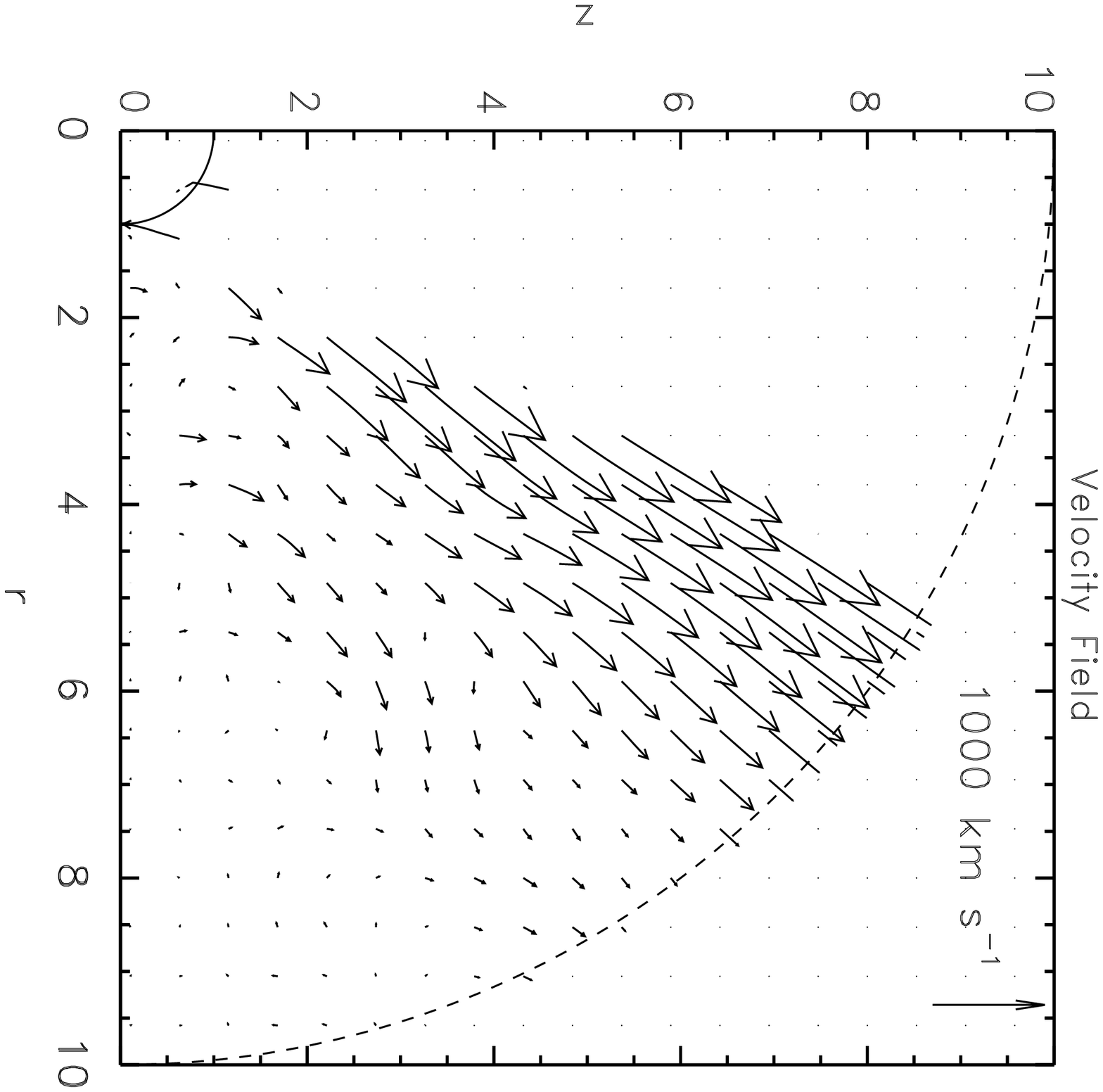}}
\put(90,0){\includegraphics{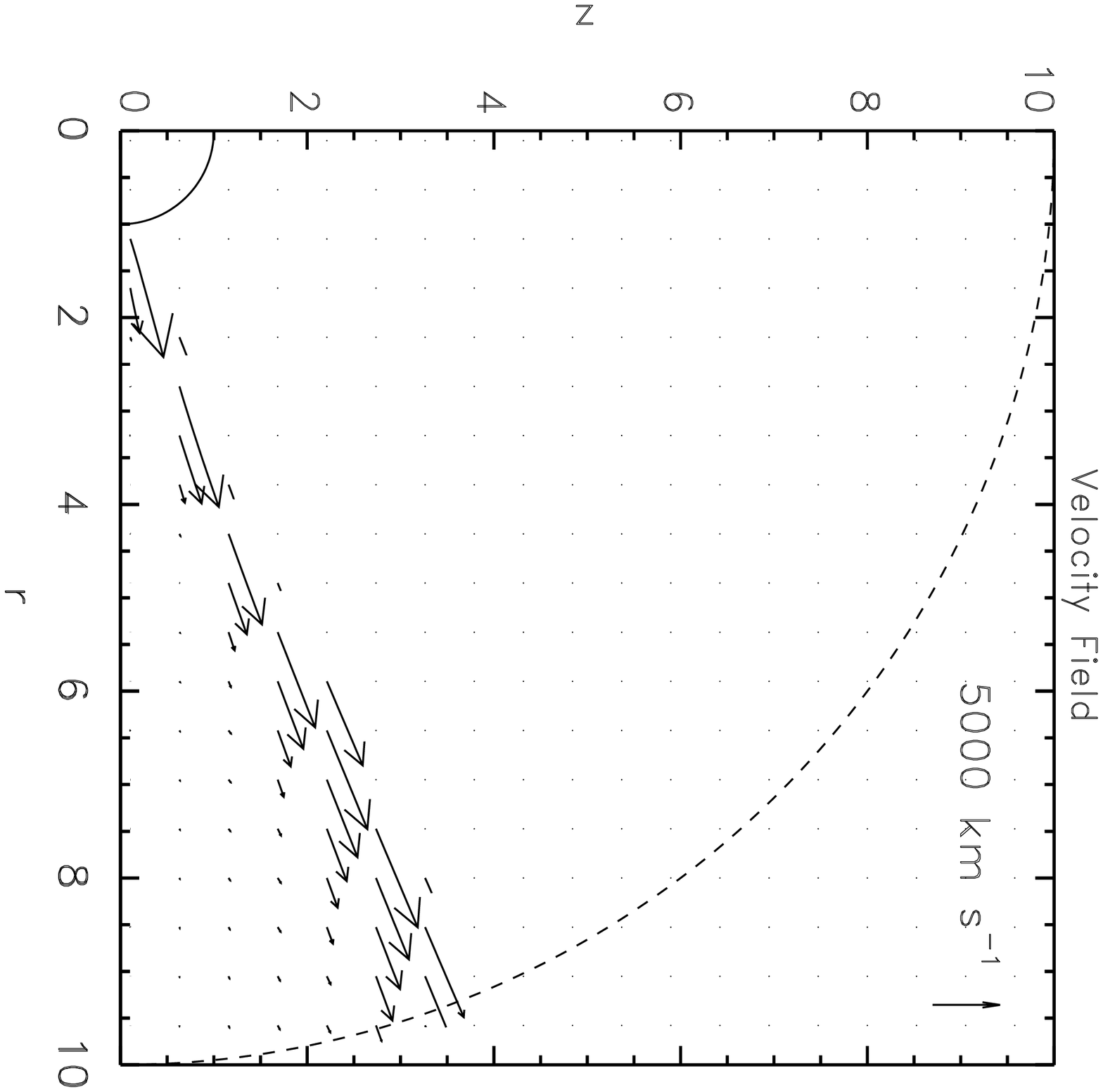}}
\put(90,210){\includegraphics{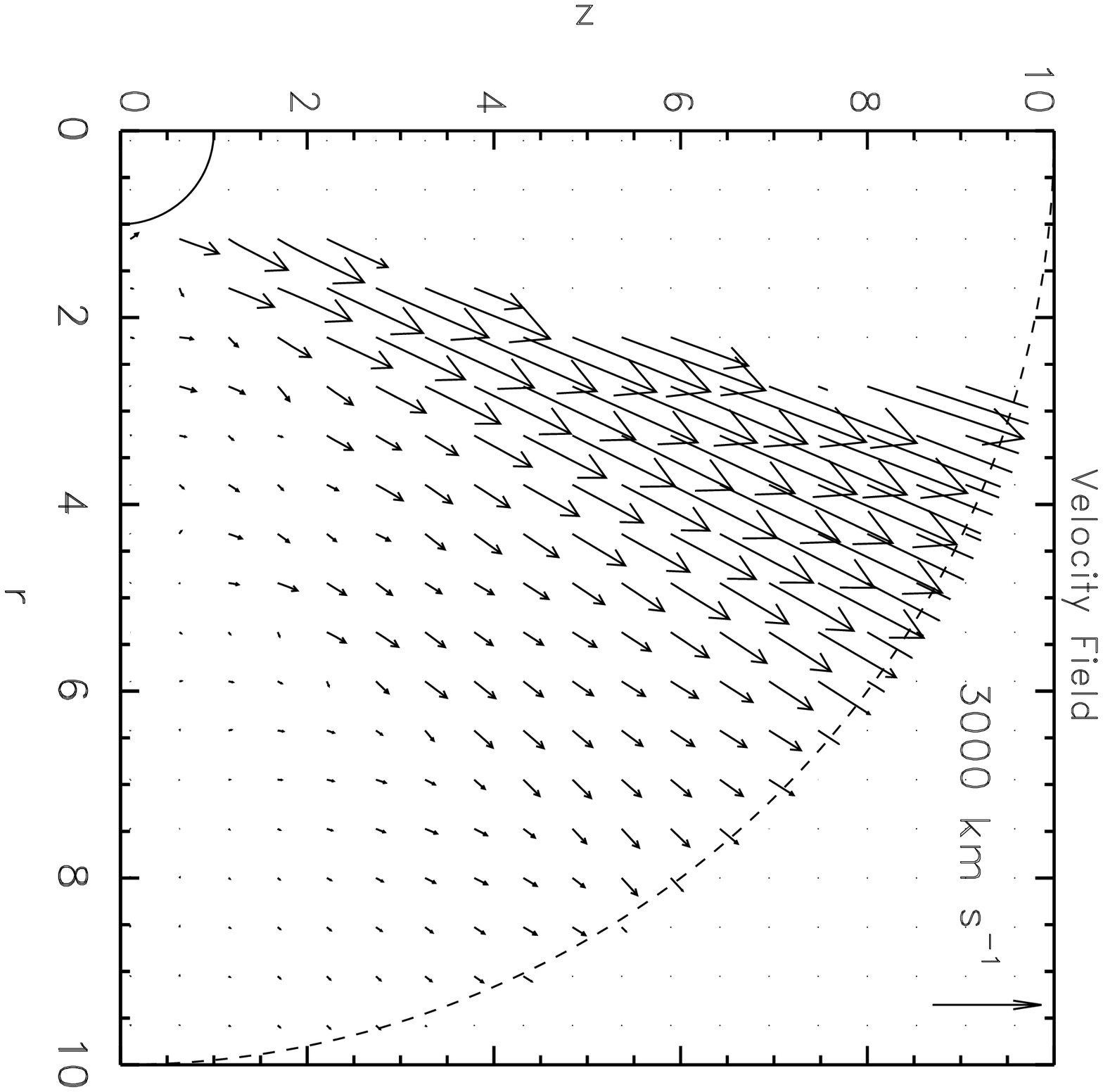}}
\end{picture}
\caption{Maps of poloidal velocity for a range of models.   
The top two panels, a and b, are both models with $x=0$ but with $\MDOT_{a} = 
10^{-8}~\MSUNYR$ (model A) and 
$\MDOT_{a} = \pi\times 10^{-8}~\MSUNYR$ (model B), respectively.  
The bottom two panels, c and d, are 
results for models both with $\MDOT_{a} = \pi \times 10^{-8}~\MSUNYR$, but
with $x=1$ (model C) and $x=3$ (model D).  The top two panels show the
effect on the outflow geometry of increasing the disk luminosity alone,
while the top right and bottom two panels show the effect of adding in an 
increasingly larger stellar component ($x$ $=$ 0, 1 and 3) to the radiation
field.  Adding in an increasingly large stellar component causes the  
outflow to become more equatorial.  Note that we suppress velocity vectors
in regions of very low density (i.e., $\rho$ less than 
$10^{-20}$~g~cm$^{-3}$).  The choice of input  parameters for the models
shown is the same as in Figure~10 of PSD.
}
\end{figure*}

\subsection{Comparison between new and previous models}

\begin{table*}
\footnotesize
\begin{center}
\caption{ Summary of results for disc winds with $\alpha=0.6$, 
$k=0.2$, and $M_{max}=4400$.}
\begin{tabular}{c c c c r c   } \\ \hline 
     &                         &   &                         &                   &            \\
 run & $\MDOT_a$               & x & $\MDOT_D$               &  $v_r(10 r_\ast)$ & $\omega$   \\ 
     & (M$_{\odot}$ yr$^{-1}$) &   & (M$_{\odot}$ yr$^{-1}$) & $(\rm km~s^{-1})$ & degrees    \\ \hline  

     &                         &   &                         &                   &             \\
our     &                         &   &                         &                   &             \\

A    &$  10^{-8}$              & 0 &  $ 5.5\times10^{-14}$     & 900                 &  50  \\
B    &$ \pi \times 10^{-8}$    & 0 &  $ 4.0\times10^{-12}$      & 3500                &  60  \\
C    &$ \pi \times 10^{-8}$    & 1 &  $ 2.1\times10^{-11}$     & 3500                &  32  \\
D    &$ \pi \times 10^{-8}$    & 3 &  $ 7.1\times10^{-11}$     & 5000                &  16  \\
E    &$ \pi \times 10^{-8}$    & 10 & $ 3.2\times10^{-10}$     & 7000                &  8  \\

      &                         &   &                         &                   &             \\

PSD's     &                         &   &                         &                   &             \\

2     &$  10^{-8}$              & 0 & $ 4.8\times10^{-14}$     & 900                 &  42  \\
3     &$ \pi \times 10^{-8}$    & 0 & $ 4.7\times10^{-12}$     & 3500                &  55  \\
8     &$ \pi \times 10^{-8}$    & 1 & $ {2.1\times10^{-11}}^{(a)}$     & 3500           &  37  \\
12    &$ \pi \times 10^{-8}$    & 3 & $ {6.3\times10^{-11}}^{(b)}$     & 5000          &  28  \\
14    &$ \pi \times 10^{-8}$    & 10 & $ {3.1\times10^{-10}} $     & 7000          &  24  \\

\hline 
\end{tabular}
\\a) We found a typographical error in PSD table 2 b) 
We calculated this model for longer than PSD did and we found that 
the flow settles at a higher mass-loss rate.
\end{center}
\end{table*}

Using our new `full-Q' method to compute the line force, we have
recomputed five models using the same parameters as runs 2, 3, 8, 12
and 14 in PSD.  This range of models illustrates the dependence of the
disk wind on the disk and stellar luminosity.  Table~1 summarises the
gross properties of our new calculations in comparison to PSD including
the mass-loss rate, $\MDOT_w$, characteristic velocity at 10$r_\ast$,
$v_r(10r_\ast)$ and flow opening angle, $\omega$.

PSD found that radiatively driven winds from disks fall into two
categories: 1) intrinsically unsteady with large fluctuations in
density and velocity, and 2) steady with smooth density and velocity
distributions.  Which type of flow is produced depends on the geometry
of the radiation field, parameterised by $x$:  the flow is unsteady if
the radiation field is dominated by the disk ($x<1$), and steady if
dominated by the star ($x \simgreat 1$).  The geometry of the radiation
field also controls the geometry of the flow; the wind becoming more
polar as $x$ decreases.  On the other hand, the mass-loss rate and
terminal velocity are insensitive to geometry and depend more on the
system luminosity, $L_D+L_\ast$.

Figure~1 compares the density in the wind in two models computed using
the method of PSD (top panels), the full-Q formalism described here but
setting the velocity gradient tensor elements which depend on the
rotational velocity to zero (middle panels), and the full-Q formalism
retaining all terms (bottom panels).  The calculations in which
$e_{r\phi}$ and $e_{\theta\phi}$ are set to zero (middle panels) are
designed to test the effect of terms related to shear in the rotational
velocity.  At the same time, this removes the azimuthal force term 
altogether.  The left
column shows the results for a model in which $\MDOT_a=10^{-8} \rm
\MSUN~yr^{-1}$ and $x=0$, the right column corresponds to
$\MDOT_a=\pi\times10^{-8} \rm \MSUN~yr^{-1}$ and $x=3$.  The former
corresponds to the fiducial unsteady wind model discussed in detail in
PSD.  The latter gives a steady wind in which the force multiplier is
well below $M_{max}$ and hence should be strongly sensitive to the velocity
gradient tensor representation.

First, we consider the changes seen in the $x=0$ wind model (left-hand
panels).  This model remains unsteady in all three treatments of the line 
force.  Details of the properties of the wind are
changed, however.  For example, the opening angle 
is increased from $42^o$ to $50^o$ between the PSD (top) and full-Q (bottom)
results.  In Table~1, it may be seen that the mass-loss rate increases by 
$\sim$20 per cent  between these two cases, while the characteristic velocity 
is the same. On the smaller scale of the flow substructure, we do find some 
subtle  differences. Most notably, the slow, dense, complex portion of the 
wind extends further above the disk plane with full-Q case as compared in PSD.

In the steady wind models (right hand panels of Figure~1), similar
fractional changes in the mass-loss rate
are observed between the PSD and full-Q formulations and, again, the
characteristic flow speeds are much the same.  
The main difference is in the wind opening angle: in the full-Q case, the 
opening angle is distinctly smaller than in PSD.  This is because the 
latitudinal component of the stellar line force gains more from the inclusion 
of the extra terms  than either the radial component, or any of the disk 
contributions.  The effect is stronger here than it happens to be in the 
unsteady wind models illustrated because the higher total luminosity
of the system ensures that the general level of the force multiplier 
remains below saturation (i.e., $M(t)~\simless~0.5~M_{max}$).

In both the steady and unsteady wind models, the impact of the rotational
shear terms is not very significant.  The comparison between the middle 
and lower panels of Figure~1 reveals that the inclusion of these terms
slightly increases the flow opening angle.  
We examined the origins of this and found that this change is due to 
enhancing various components of the line force mainly in the inner
disk.

In summary, our finding that the mass-loss rate and velocity are similar in 
the approximate PSD and full-Q cases can be explained by the fact that 
the terms in the velocity gradient tensor used by PSD are indeed dominant.

\subsection{Overview of trends in the full-Q models}

Figure 2 plots the poloidal velocity in four models computed with 
the full-Q method.  Panels a and b compare the flow pattern from two models,
with $x = 0$
in which the mass accretion rate and hence the disk luminosity $L_D$
is increased: specifically, $\MDOT_a$ is raised from 
$10^{-8} \rm \MSUN~yr^{-1}$ to $\pi\times 10^{-8} \rm \MSUN~yr^{-1}$.
By contrast, panels b, c, and d compare the flow pattern from
three models in which the mass accretion rate is held fixed at 
$\MDOT_a=10^{-8} \rm \MSUN~yr^{-1}$ while the stellar
luminosity is varied using, $x=0$, 1, and 3.  This diagram presents models
with the same input parameters as those shown
in Figure 10 in PSD, with the difference that here full-Q
is implemented in their calculation and we plot poloidal velocity vectors 
instead of density.

Our new models confirm PSD's result that the flow becomes more 
equatorial as the contribution of the central star to the radiation field 
increases. However the scale of the changes in the flow geometry is greater for
the full-Q case than in the approximate Q case -- the models for low $x$ are 
more polar here than in PSD, whereas the models for high $x$ are more 
equatorial.  PSD found that the reduction of the opening angle of the disk 
wind slows appreciably for $x\simgreat 3$.  With full-Q, this slowing is
deferred until $x \simgreat 5$.  For example, we calculated the model for 
$x=10$ and found $\omega=8^o$, rather than $\omega=24^o$ for PSD's 
corresponding model, run~14.  Despite this geometric change, the gross
wind properties as listed in Table 1 are scarcely any different.

The two models illustrated in Figure 1 showed that the model presented
in PSD remains unsteady when recalculated using full Q, and that the
originally steady model is still steady.  We can generalise this further
in that we find no noticeable shift in the value of $x$ ($= L_{\ast}/L_D$)
at which the change from unsteady to steady occurs.  This is a further
respect, to add to the mass-loss rate and characteristic flow speed, 
in which the full-Q models continue to closely resemble PSD's  models.

\section{Discussion and Summary}

The efficient algorithm described here has allowed us to examine
the effects of all terms in the velocity gradient tensor
on the structure of line-driven winds from disks.  We find that
the qualitative features of such winds are not changed by the more
accurate algorithm used here.  In particular, models which displayed
unsteady behavior in PSD are also unsteady with the full-Q method.
This indicates the approximations adopted in PSD indeed captured
the dominant terms in the line force.

On generalizing the line force, we determine the geometry and strength of
the line force in an exact way for a constant geometry of the radiation 
field.  We continue to find, as in PSD, that the mass-loss rate and 
characteristic velocity do not depend on either of these two geometries but 
primarily on the total system luminosity. This is in keeping with the 
conclusion reached by Proga (1999) who showed that the mass-loss rate of even 
a simple spherically-symmetric stellar wind is of the same order of
magnitude as that of a pure disk wind of the same total luminosity.

The dependence of the disk mass-loss rate on the total system luminosity, 
$L_D+L_\ast$, indicates that the irradiation due to the central star can power 
disk mass loss as does the disk radiation.  Indeed we have already shown 
that the radiation from the luminous central star can  drive a wind from 
an optically-thick disk of negligible intrinsic luminosity, i.e., for $x=300$ 
and $L_D M_{max} << L_{Edd}$, where $L_{Edd}=4\pi G M_\ast/\sigma_e$ is the Eddington luminosity 
(Drew, Proga \& Stone 1998).  The significance of 
irradiation has also been studied by Gayley, Owocki \& Cranmer (1999).  
They also find, on the basis of a quite different formulation of the problem 
for an irradiated planar slab atmosphere, that the irradiation enhances or 
even induces the mass loss.

Whilst the radiation field geometry typically controls the geometry of 
the flow in the way described here and in PSD, there is a complicating effect 
that comes into play at low luminosities that has so far gone unremarked.  
When $L_D M_{max} \simless$~a~few~$L_{Edd}$, we find that 
the upper bound, $M_{max}$ on the force multiplier exerts an influence on 
the geometry of the disk wind. In particular for $x=0$, our models show 
that the higher $M_{max}$, the higher the wind opening angle becomes. 
Proga (1999) calculated a few models for $x=0$ without saturation of $M(t)$ 
(his models in the CAK\&FD case) and found that $\omega \approx 90^o$ 
regardless of disk luminosity. Calculation using our full-Q method confirm
this result.

The changes caused by the full-Q treatment in disk winds are not so serious
as in winds around rapidly rotating oblate stars. OCG's treatment of the
rotating stellar wind case was a big step forward in that their
inclusion of all terms in Q introduced latitudinal and azimuthal components
of the line force where, before, only the radial component had been
considered. They found that (i)
the latitudinal component is poleward and unopposed by any other force and 
hence is dynamically significant in inhibiting the equatorward drift of 
the wind; (ii) the azimuthal line force
acts against the sense of rotation and is less significant because
it causes only a modest spin-down of the wind rotation. 
In the disk wind case, however, there is only one qualitative change with 
respect to our own earlier treatment (PSD) when all Q terms are included: 
there is now a non-zero azimuthal line force.  Again, as in the rotating
stellar wind case, this
is rather  weak as compared to the other components and is not of great 
importance. 
The importance of this azimuthal term is further weakened by the fact that
it can spin-down or spin-up the wind rotation depending on
location and time. Its sign may change in disk 
winds because  the wind velocity field is complex and so the contribution of 
all  Q terms symmetric in $\phi$ can be positive or negative
(even $dv_r/dr$ is negative in some regions of disk winds).

The fact that the unsteady behavior observed in our models
has not changed with a more accurate treatment of the radiation
force indicates it is indeed a robust property of line driven winds
from disks.  Why does increasing the radial component of the
radiation force 'organize' the wind into a steady state?
Let $r'$ and $z'$ define position along a streamline in the
wind in cylindrical coordinates.
An increase of the vertical component of the gravity, 
\begin{equation}
g_z \propto -\frac{z'}{(r'^2+z'^2)^{3/2}}
\end{equation}
with height at a fixed radius $r'$ is the main driver of the unsteady flow. 
However this increase of the gravity can be significantly reduced
if the streamlines are directed outwards from purely vertical (i.e., $r'$
increases with $z'$).  At the same time, this tilt also brings into play
an increase of the horizontal effective gravity, $g_r$, along each streamline:
\begin{equation}
g_r \propto \frac{r_f}{r'^3}-\frac{r'}{(r'^2+z'^2)^{3/2}},
\end{equation}
where $r_f$ is the radius on a Keplerian disk at which a streamline
originates.  However the increase of $g_r$ with $r'$ is slower than the
increase of $g_z$ with $z'$ because of the decaying centrifugal term.
In other words, the line force can more easily maintain domination
over gravity if the flow climbs the gentler gravitational hill in the
horizontal direction as compared with the vertical direction.
Furthermore,
driving material along  streamlines outward from the vertical causes
density to decline as $1/r'$ as required by geometrical dilution -- this,
very usefully, tends towards increasing the line force, thereby 
facilitating a better match with trends in gravity.

Despite our progress in developing realistic models for line-driven disk
winds, limitations obviously  remain. For example we calculate 
the line-force using the Sobolev approximation. It is questionable if this 
approximation is valid in cases where the wind is slow and/or highly
structured. Even if the Sobolev approximation is locally valid in those 
cases, our full-Q method 
does not account for non-local effects in line transfer.  For example, 
our line force is potentially overestimated in upper parts of the flow 
because of neglect of self-shadowing. Our models also assume axisymmetry --
in particular that the disk plane is flat and perpendicular to the 
rotational axis whereas in reality disks may be tilted or twisted.
We also simplify
the thermal structure of the disk wind by assuming that it is isothermal.
In our models, the disk is not isothermal to start with, and we need also to 
calculate of the ionization structure and energy balance to properly
model the wind thermodynamics and the radiation pressure.  

Another aspect of our models to date that will need to be revisited in
future is the prescription for the reemission of disk irradiation.  We have
assumed thus far that the disk re-emits all absorbed energy locally
and isotropically.  Gayley et al. (1999) have opted for more or less
the opposite prescription appropriate to purely scattering 
atmospheres.  Strictly, detailed NLTE photoionization calculations of 
externally-irradiated disks need to be performed on a case-by-case basis as 
it is not obvious a priori what prescription should be used in
any one situation.

Models of line-driven winds from luminous disks show promise of being
able to explain mass loss phenomena associated with e.g., cataclysmic 
variables, massive young stellar objects and extreme Be objects (e.g., PSD ,
Drew, Proga \& Stone 1998, Oudmaijer et al. 1998). 
Our further progress in understanding those winds requires a direct, 
quantitative comparison between observations and model predictions. 
Therefore we plan to use our three-dimensional disk wind structure to calculate
synthetic line profiles -- initial work is underway. We also plan to further 
develop our disk wind models. For example, we intend to calculate the 
photoionization structure
of the wind that will allow  to take into account changes of the line force
with local ionization. With such improvements our models will be readily
applicable to the systems with a very wide dynamical range such as AGN.
An outcome of the present study has been to confirm the unsteady behaviour 
in predominantly disk-illuminated models, first identified in the more
approximate calculations due to PSD.  To test this, we have obtained an
allocation of Hubble Space Telescope time to carry out a high time- and
spectral-resolution study of 3 nova-like variables -- objects in which we
expect disk winds to be driven primarily by disk radiation.  If winds in 
these systems are unsteady then we hope to see evidence of this in the form
of time-variable fine structure in blueshifted absorption features.

{\bf Acknowledgments:}
This research has been supported by a research grant from PPARC.  Computations
were performed at the  Imperial College Parallel Computing Centre.



\begin{thebibliography}{}
\bibitem[]{}
  Batchelor G.K. 1967, An Introduction to Fluid Mechanics (Cambridge:
Cambridge University Press)
\bibitem[\protect\citename{CAK}1975]{CAK} 
  Castor J.I., Abbott D.C.,  Klein R.I., 1975, ApJ, 195, 157 (CAK)
\bibitem[]{}
  Drew J.E., Proga D., Stone J.M., 1998, MNRAS, 296, L6 
\bibitem[\protect\citename{Friend \& Abbott}1986]{fa}
  Friend D.B., Abbott D.C., 1986, ApJ, 311, 701
\bibitem[]{}
 Gayley K.G., Owocki, S.P., Cranmer, S.R. 1999, ApJ, 513, 442
\bibitem[\protect\citename{Icke}1980]{ic0}
  Icke V., 1980, AJ, 85, 329
\bibitem[]{}
  Murray N., Chiang J., Grossman S.A. Voit, G.M., 1995, ApJ, 451, 498
\bibitem[]{}
  Oudmaijer R. D., Proga D., Drew J. E., de Winter D., 1998, MNRAS, 300,
  170
\bibitem[\protect\citename{Owocki et al. }1988]{OCR}
  Owocki S.P., Castor J.I., Rybicki, G.B. 1988, ApJ, 335, 914
\bibitem[]{}
 Owocki, S.P., Cranmer, S.R.,  Gayley K.G.  1996, ApJ, 472, L115 (OCG)
\bibitem[\protect\citename{Pauldrach et al. }1986]{ppk}
  Pauldrach A., Puls J.,  Kudritzki R.P., 1986, A\&A, 164, 86 
\bibitem[\protect\citename{Pereyra et al. }1997]{pe}
  Pereyra N.A., Kallman T.R.,  Blondin J.M., 1997, ApJ, 477, 368
\bibitem[]{}
  Proga D. 1999, MNRAS, 304, 938
\bibitem[]{}
  Proga D., Stone J.M., Drew J.E., 1998, MNRAS, 295, 595 (PSD)
\bibitem[\protect\citename{Rybicki \& Hummer}1978]{rh78} 
  Rybicki G.B.,  Hummer D.G.  1978, ApJ, 219, 654
\bibitem[\protect\citename{Shakura \& Sunyaev}1973]{ss} 
  Shakura N.I., Sunyaev R.A. 1973 A\&A, 24, 337
\bibitem[\protect\citename{Stone \& Norman}1992]{sn92}
  Stone J.M., Norman M.L. 1992, ApJS, 80, 753
\bibitem[\protect\citename{Vitello \& Shlosman}1988]{vs} 
  Vitello P.A.J.,  Shlosman I., 1988, ApJ, 327, 680
\end{thebibliography}
\end{document}